%% file: hoffmann2.tex
\documentclass[graybox]{svmult}

\usepackage{graphicx}
 \usepackage[cmex10]{amsmath}
 \usepackage{amsfonts}
\usepackage{amssymb}
\usepackage{cite}
\usepackage{mathptmx}       
\usepackage{helvet}         
\usepackage{courier}        
\usepackage{type1cm}        
\usepackage{makeidx}         
\usepackage{graphicx}        
\usepackage{multicol}        
\usepackage[bottom]{footmisc}

\usepackage[normalem]{ulem} 

\makeindex             

\begin{document}

\title*{Random Walks on Stochastic Temporal Networks}


\author{Till Hoffmann, Mason A. Porter, and Renaud Lambiotte}

\institute{Till Hoffmann \at Department of Physics, University of Oxford, \email{tillahoffmann@gmail.com} \and Mason A. Porter \at Oxford Centre for Industrial and Applied Mathematics, Mathematical Institute, University of Oxford and CABDyN Complexity Centre, University of Oxford, \email{porterm@maths.ox.ac.uk} \and Renaud Lambiotte \at Department of Mathematics, University of Namur, \email{renaud.lambiotte@fundp.ac.be}}


%
%
\maketitle

\abstract*{In the study of dynamical processes on networks, there has been intense focus on network structure---i.e., the arrangement of edges and their associated weights---but the effects of the temporal patterns of edges remains poorly understood. In this chapter, we develop a mathematical framework for random walks on temporal networks using an approach that provides a compromise between abstract but unrealistic models and data-driven but non-mathematical approaches. To do this, we introduce a stochastic model for temporal networks in which we summarize the temporal and structural organization of a system using a matrix of waiting-time distributions. We show that random walks on stochastic temporal networks can be described exactly by an integro-differential master equation and derive an analytical expression for its asymptotic steady state. We also discuss how our work might be useful to help build centrality measures for temporal networks.
}

\abstract{In the study of dynamical processes on networks, there has been intense focus on network structure---i.e., the arrangement of edges and their associated weights---but the effects of the temporal patterns of edges remains poorly understood. In this chapter, we develop a mathematical framework for random walks on temporal networks using an approach that provides a compromise between abstract but unrealistic models and data-driven but non-mathematical approaches. To do this, we introduce a stochastic model for temporal networks in which we summarize the temporal and structural organization of a system using a matrix of waiting-time distributions. We show that random walks on stochastic temporal networks can be described exactly by an integro-differential master equation and derive an analytical expression for its asymptotic steady state. We also discuss how our work might be useful to help build centrality measures for temporal networks.
}



\section{Introduction}
\label{sec:1}

A broad variety of systems are composed of interacting elements (e.g., nodes connected by edges) and can be represented as networks. Important examples include the Internet, highways and other transportation systems, and many social and biological systems.  Because of the ubiquity of network representations, the study of networks has emerged as one of the fundamental building blocks in the study of complex systems \cite{bocca,evans,review}.  

Unfortunately, most empirical studies of networks thus far have been based on observations of snapshots of systems. Similarly, most theoretical efforts have focused on static network properties (such as degree distribution or modularity) and their effects on dynamical processes such as diffusion or epidemic spreading. In its current state, network theory thus fails to properly model a broad range of complex systems, in which interactions need not be active permanently but rather might switch on only temporarily. Observations in financial markets \cite{sabatelli2002waiting}; email \cite{burstystreams,Eckmann,bara,malmgren}, letter \cite{oliveira}, and online \cite{burstystreams2,inform} communication networks; face-to-face contacts \cite{isella}; movie rentals \cite{mariano}; and many other situations have illustrated that the time intervals between edge activations can be distributed nontrivially, and the ensuing dynamics thus tend to deviate significantly from Poisson processes. It has also been shown that such burstiness yields radically different dynamics on networks, thereby making models of dynamical processes on static networks inappropriate in many situations \cite{importanceofdynamics0,importanceofdynamics,vazquez2,Moro,tang2010,rocha,karsai,caley2007influenza,takaguchi,miritello,maxi}.

\begin{figure} 
	\centerline{\includegraphics[width=0.95\textwidth]{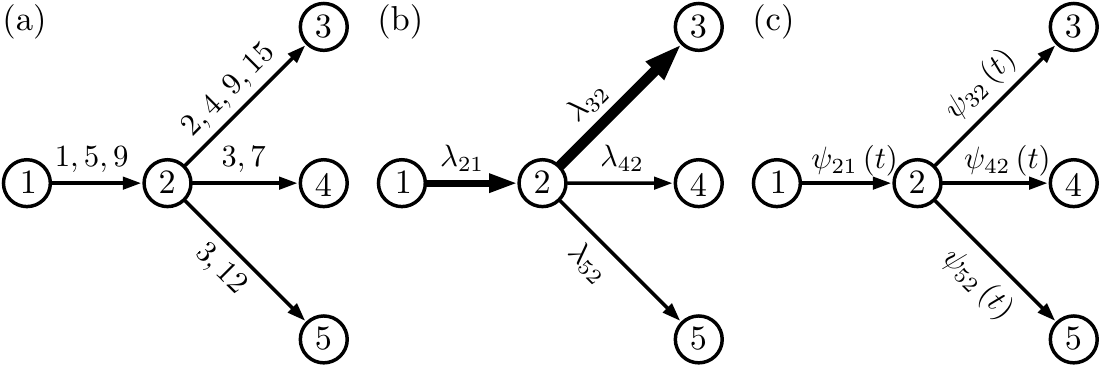}}
	\caption{\label{fig:ss-example} 
Different levels of abstraction to model dynamics on temporal networks. In this illustration, a walker steps from node $1$ to node $2$ and then jumps to one of its three neighbors. When studying dynamics on temporal networks, researchers tend to either (a) perform numerical simulations on empirical networks using the observed times at which edges are active between nodes or (b) develop mathematically-tractable Markovian models that neglect temporal patterns (typically by considering only aggregations of the interaction between nodes). In this chapter, we aim at finding (c) an intermediate level of modelling, in which we replace the sequence of activation times by a stochastic model that preserves a system's inter-event distribution.  We thereby balance the amount of data included in the description with the description's simplicity.
} 
\end{figure}

A natural framework to study many time-dependent complex systems is to use temporal networks  \cite{reviewSaram}, in which one accounts for the timings of interactions instead of assuming static connectivity (e.g., by employing data aggregation) or that interactions take place at a uniform rate.  Two approaches (see Fig.~\ref{fig:ss-example} ) have been used to add a temporal dimension to networks to account for constraints imposed by temporality on spreading processes. First, one can perform simulations on temporal graphs for which a time series of the presence versus absence of edges is deduced directly from empirical observations \cite{rocha,karsai}. However, such a computational approach has a significant drawback, as it relies entirely on numerical simulations and is unable to provide a general picture of such problems. Second, one can use an abstract approach by developing spreading models that nevertheless attempt to incorporate realistic temporal statistics. Such models can then be studied either mathematically or using numerical simulations \cite{takaguchi,vazquez2}. In this second approach, an underlying network is studied as a fluctuating entity that is typically driven by a stationary stochastic process. This approach is nice because it is amenable to mathematical analysis \cite{new1,Moro,new2,miritello,Iri}, and it also provides a more accurate picture of time-dependent complex systems than do static networks.

In this paper, which is a modified version of Ref.~\cite{hoffmann}, we take the second approach and develop a mathematical framework to explore the effect of non-Poisson inter-event statistics on random walks. To do this, we apply the concept of a generalized master equation \cite{montroll1965}, which is traditionally defined on regular lattices, to the study of continuous-time random walks on arbitrary networks. Generalized master equations are a standard tool in non-equilibrium statistical physics; they lie at the heart of the theory for anomalous diffusion, and their applications range from ecology \cite{ap2} to transport in materials \cite{ap1}. Our choice regarding what dynamics to consider is motivated by the importance of random walks as a way to understand how network structure affects dynamics and to uncover prominent structural features from networks.

The rest of this paper is organized as follows. We first introduce basic concepts of random walks on static networks and then introduce a model for stochastic temporal networks. In Section \ref{sec3}, we derive a generalized master equation to describe random walks on stochastic temporal networks. We examine the stationary solution of this equation and show that it is determined by an effective transition matrix whose dominant eigenvector can be calculated rapidly even in very large networks if they are sparse. After checking that the generalized master equation reduces to standard rate equations when the underlying process satisfies Poisson statistics, we validate theoretical predictions using numerical simulations. Finally, we discuss the implications of our work for constructing centrality measures of nodes in temporal networks.


\section{Random Walks on Static Networks}\label{sec2}

In this section, we review basic properties of random walks on static networks. The structure of a network is described by an $N \times N$ adjacency matrix $A$, where $N$ is the number of nodes in the system. By definition, the adjacency matrix component $A_{ij}$ gives the weight of an edge going from $j$ to $i$. The adjacency matrix reflects the underlying network structure, on which various dynamical processes (e.g., diffusion) can occur. The simplest process that can be defined on a network is a discrete-time, unbiased random walk.  At a given step of such a process, a walker located at a node $j$ follows a given edge leaving $j$ with a probability proportional to the edge's weight. 

The expected density $n_{i;t}$ of walkers at node $i$ at time $t$ evolves according to the evolution equation 
\begin{equation}\label{discrete}
	n_{i;t+1} = \sum_{j} T_{ij} n_{j;t}\,,
\end{equation}
where $T_{ij}$, a component of the transition matrix $T$, represents the probability to jump from $j$ to $i$ and is defined as
\begin{equation} \label{unbiased}
	T_{ij}=A_{ij}/s_{j}^{\rm out} \,,
\end{equation}
where the out-strength 
\[ 
	s_j^{\rm out} = \sum_i A_{ij}
\]
 of node $j$ is the total weight of the edges leaving node $j$. Because the total number of walkers is conserved, the columns of $T_{ij}$ are normalized: 
\[ 
	\sum_{i} T_{ij}=\frac{1}{s_j^{\rm out}} \sum_i A_{ij}=1\,. 
\]
The condition $\sum_i n_{i;t}=1$ is thus verified for all times $t$. The dynamical process (\ref{discrete}) with transition matrix components (\ref{unbiased}) has been studied in detail, and a wide variety of its properties are known for many different types of networks \cite{chung0}. For instance, consider a network that is undirected, which implies that $A_{ij}=A_{ji}$ and that the strength of node $j$ is $s_i = s_i^{\rm out}$, and also suppose that it is connected and non-bipartite.  The above discrete-time random walk then converges to a unique equilibrium solution $p$ with components
\begin{equation}\label{classicalsol}
	p_i=s_i/W\,, 
\end{equation}
where $W=\sum_i s_i$.

By construction, $p$ is the dominant eigenvector (i.e., the eigenvector corresponding to the maximum positive eigenvalue) of the transition matrix.  Its corresponding eigenvalue is 1, so it satisfies
\[ 
	\sum_{j} T_{ij} p_j=p_i\,. 
\]

When modelling diffusion, it is often desirable (and more realistic) to allow walkers to jump in an asynchronous fashion. A natural way to implement such a situation is to switch from a discrete-time to a continuous-time perspective \cite{lambiotte}. One then needs to use a so-called \emph{waiting-time distribution} (WTD), which determines the time that is spent by a walker on a node before traversing one of the available edges. The most common assumption is to use an exponential WTD:
\[ 
	\psi(i;t)=\lambda_i e^{- \lambda_i t}\,,
\]
for which the process is Markovian. The rates $\lambda_i$ at which a walker jumps can in general be non-identical and depend on the node $i$ on which the walker is located. Such continuous-time random walks are governed by the differential equation
\begin{equation}\label{ctrwgeneral}
	\frac{d{n}_{i}}{dt} = \sum_{j} \left( \frac{A_{ij}}{s_j} \lambda_j  - \lambda_i \delta_{ij}\right) n_j \equiv - \sum_{j} L_{ij} n_j\,,
\end{equation}
where $L_{ij}$ is the component of the Laplacian matrix describing the dynamics. For undirected networks, the stationary solution has components 
\begin{equation}\label{solgeneral}
	p_i=\frac{s_i}{Z \lambda_i}\,, 
\end{equation}
where $Z=\sum_i s_i/\lambda_i$ is a normalization constant. One can interpret the quantity $p_i$ as the frequency at which a node $i$ is visited multiplied by the characteristic time $ \langle t_i \rangle\approx1/\lambda_i$ spent on it.  This frequency is proportional to $s_i$, which is the same as for discrete-time random walks (\ref{classicalsol}).  Standard choices for the jumping rates include the uniform rate $\lambda_i=1$, for which
\begin{equation}\label{l1} 
	L_{ij}=\delta_{ij}-A_{ij}/s_j\,,
\end{equation} 
and a rate $\lambda_i=s_i$ proportional to node strength, for which
\begin{equation}\label{l2}
	L_{ij}= s_i \delta_{ij} - A_{ij}\,.
\end{equation}
These are the two standard types of graph Laplacians.  With the choice (\ref{l2}), the steady-state solution of a Poisson random walk is uniform (i.e., $p_i=1/N$, which is independent of $A$) regardless of the topology and edge weights of a network.

\section{Stochastic Temporal Networks} \label{sec3}

The models that we described in Section \ref{sec2} overlook temporal patterns of the activation of edges by assigning a single scalar $A_{ij}$ to represent an aggregation of the activity between nodes $i$ and $j$. Such an aggregate measure of the importance of the connection between $i$ and $j$ is often understood as the rate at which edges are selected by a walker, and one can certainly use this perspective in equation (\ref{ctrwgeneral}). The probability for an edge to be selected in a time interval $dt$ is thus independent of the time that has elapsed since the process started \cite{Hoel1971Introduction}. This Markovian assumption facilitates theoretical analysis, but it is accurate only for systems in which the rate at which events take place is not history-dependent.

To go beyond a Markovian picture, we propose to study random walks on networks that evolve in time in a stochastic manner. We assign to each edge an inter-event time distribution that determines when an edge is accessible for transport. The dynamics of a network are thus characterized by an $N\times N$ matrix $\psi\left(t\right)$ of waiting-time distributions $\psi_{ij}\left(t\right)$ that determine the appearance of an edge emanating from node $j$ and arriving at node $i$. We also assume that the edges remain present for infinitesimally small times.  That is, no network edge is present except at random instantaneous times determined by $\psi\left(t\right)$, when a single edge is present. From now on, we employ the following terminology. We use $\mathcal{G}$ to denote an underlying graph that determines which edges are allowed and which are not, and $\psi_{ij}\left(t\right)$ determines a dynamic graph in which edges appear randomly according to the assigned waiting times. We use such a random process to model the transitions of a random walker moving on $\mathcal{G}$. By construction, a walker located at a node $j$ remains on it until an edge leaving $j$ toward some node $i$ appears. When such an event occurs, the walker jumps to $i$ without delay and then waits until an edge leaving $i$ appears.

There are several ways that one can set up the WTDs $\psi_{ij}\left(t\right)$. One option is to consider the processes leading to steps being associated with an \emph{active} walker. In this case, the walker's clock is reinitialized when it makes a step to a node. For example, gossip traversing a social network might be modeled using an active walker because the process of broadcasting gossip restarts when the gossip is received by a new contact. Another option is to consider the processes leading to steps being associated with a \emph{passive} walker. In this case, the edges' clocks are reinitialized when they are activated. A virus spreading across a social network might be modeled using a passive walker because interactions among people are not (primarily) initiated by a virus. In the present work, we consider the case of an active walker in which a WTD corresponds to the probability for an edge to occur between times $t$ and $t+dt$ after the random walker arrives on node $j$ in the previous step.


It follows from the definition of the WTDs as probability distribution functions that
\[ 
	\int_{0}^{\infty} \psi_{ij}\left(t\right)=1\,. 
\] 
The probability that an edge appears between $j$ and $i$ before time $t$ is 
\[
	\int_{0}^{t}\psi_{ij}\left(t'\right)dt'\,, 
\] 
and the probability that it does not appear before time $t$ is therefore
\begin{equation} \label{eq:chi}
	\chi_{ij}\left(t\right)\equiv1-\int_{0}^{t}\psi_{ij}\left(t'\right)dt'\,. 
\end{equation}
If a transition from $j$ to $i$ is not allowed, then the corresponding element $\psi_{ij}(t)$ is equal to $0$ for all times. 

It is important to distinguish between the WTD $\psi_{ij}\left(t\right)$ of the process that might lead to a step along a network and the probability distribution $T_{ij}\left(t\right)$ for actually making a step from $j$ to $i$. This distinction is necessary because all of the processes on a node are assumed to be independent of one another, but the probability to make a step depends on all of the processes. As an illustration, consider a walker on a node $j$ with only one outgoing edge to $i$. The probability distribution function (PDF) to make a step to $i$ a time $t$ after having arrived on $j$ is then 
\[
	T_{ij}\left(t\right)=\psi_{ij}\left(t\right)\,. 
\] 
However, if there exists another edge leaving $j$ (e.g., an edge to node $k$), then the PDF to make a transition to $i$ is modified, as a step to $i$ only occurs if the edge to $i$ appears before the one going to $k$. In this situation, 
\[
	T_{ij}\left(t\right)=\psi_{ij}\left(t\right)\chi_{kj}\left(t\right)\,. 
\] 

In general, the PDF to make a step from $j$ to $i$ accounting for all other processes on $j$ is
\begin{align} 
	T_{ij}\left(t\right) & =\psi_{ij}\left(t\right)\times\prod_{k\neq i}\chi_{kj}\left(t\right) \nonumber \\ 
& =\psi_{ij}\left(t\right)\times\prod_{k\neq i}\left(1-\int_{0}^{t}\psi_{kj}\left(t'\right)dt'\right)\,.
	\label{eq:T_ij(t)} 
\end{align} 
Equation (\ref{eq:T_ij(t)}) emphasizes the importance of the temporal ordering of the edges in the random walk. In particular, it gives greater importance to edges that tend to appear before others and are thus more likely to be selected by a random walker. 


\section{Generalized Montroll-Weiss Equation}

We now focus on the trajectories of a random walker exploring the stochastic temporal network that we described in Section \ref{sec3}. We closely follow the standard derivation of the Montroll-Weiss (MW) equation \cite{montroll1965}, which is traditionally defined on regular lattices, and we generalize it to an arbitrary $N$-node network of transitions. We are interested in finding the probability $n_{i}\left(t\right)$ to find a walker on node $i$ at time $t$.  This probability is given by the integral over all probabilities $q_{i}\left(t'\right)$ of having arrived on node $i$ at time $t'<t$, weighted by the probability $\phi_{i}\left(t-t'\right)$ of not having left the node since then:
\begin{equation}\label{convoluted}
	n_{i}\left(t\right)=\int_{0}^{t}\phi_{i}\left(t-t'\right)q_{i}\left(t'\right)dt'\,.
\end{equation} 
Taking the Laplace transform 
\begin{equation*}
	\hat{n}_{i}\left(s\right)\equiv\mathcal{L}\left\{ n_{i}\left(t\right)\right\}=\int_{0}^{\infty}n_{i}\left(t\right)e^{-st}dt
\end{equation*} 
allows us to exploit the fact that the convolution in (\ref{convoluted}) reduces to a product in Laplace space:
\begin{equation}
	\hat{n}_{i}\left(s\right)=\hat{\phi}_{i}\left(s\right)\hat{q}_{i}\left(s\right)\,.
	\label{eq:n_i(s)}
\end{equation}

We obtain the quantity $\hat{\phi}_{i}\left(s\right)$ in (\ref{eq:n_i(s)}) as follows. The probability distribution to make a step from node $i$ to any other node is
\begin{equation}
	T_{i}\left(t\right)=\sum_{j=1}^{N}T_{ji}\left(t\right)\,. 
	\label{eq:T_i(t)}
\end{equation} 
The PDF to remain immobile on node $i$ for a time $t$ is thus given by
\begin{align*} 
	\phi_{i} & =1-\int_{0}^{t}T_{i}\left(t'\right)dt'\,,
\end{align*} 
whose Laplace transform is
\begin{equation}
	\hat{\phi}_{i}\left(s\right)=\frac{1}{s}\left(1-\hat{T}_{i}\left(s\right)\right)\,.
	\label{eq:phi_i(s)} 
\end{equation}

The quantity $\hat{q}_{i}\left(s\right)$ is the Laplace transform of the PDF $q_{i}\left(t\right)$, which describes the probability of arriving on node $i$ exactly at time $t$. One calculates it by accounting for all $k$-step processes that can lead to such an event \cite{scher1973}: 
\[
	q_{i}\left(t\right)\equiv\sum_{k=0}^{\infty}q_{i}^{\left(k\right)}\left(t\right)\,,
\] 
where $q_{i}^{\left(k\right)}\left(t\right)$ represents the probability to arrive on node $i$ at time $t$ in exactly $k$ steps. Note that the PDF at node $i$ is related to that at node $j$ by the recursion relation 
\begin{equation}\label{recur} 
	q_{i}^{\left(k+1\right)}\left(t\right)
	=\int_{0}^{t}d\tau\sum_{j}T_{ij}\left(t-\tau\right)q_{j}^{\left(k\right)}\left(\tau\right)\,.
\end{equation} 
In other words, the probability to arrive on node $i$ in $k+1$ steps is the probability to arrive at any other node $j$ in $k$ steps weighted by the probability of making a step $j\rightarrow i$ at the required time. Upon taking a Laplace transform, equation (\ref{recur}) becomes 
\[
	\hat{q}_{i}^{\left(k+1\right)}\left(s\right)=\sum_{j}\hat{T}_{ij}\left(s\right)\hat{q}_{j}^{\left(k\right)}\left(s\right)\,.
\] 
Summing over all $k$ and adding $\hat{q}_{i}^{\left(0\right)}\left(s\right)$ yields 
\begin{align*}
	\hat{q}_{i}^{\left(0\right)}\left(s\right)+\sum_{k=0}^{\infty}\hat{q}_{i}^{\left(k+1\right)}\left(s\right)
	&=\sum_{j}\hat{T}_{ij}\left(s\right)\sum_{k=0}^{\infty}\hat{q}_{j}^{\left(k\right)}\left(s\right)+\hat{q}_{i}^{\left(0\right)}\left(s\right)\,,
\end{align*} 
which can also be written in terms of matrices and vectors as
\begin{align*}
	\hat{{q}}\left(s\right) 
	&=\hat{T}\left(s\right)\hat{{q}}\left(s\right)+\hat{{q}}^{\left(0\right)}\left(s\right)\,.
\end{align*} 
Noting that ${q}^{\left(0\right)}\left(t\right)={n}\left(0\right)\delta\left(t\right)$, we see that the last term is simply ${n}\left(0\right)$, which leads to the following solution in Laplace space: 
\begin{equation}
	\hat{{q}}\left(s\right)=\left(I-\hat{T}\left(s\right)\right)^{-1}{n}\left(0\right)\,.
	\label{eq:q(s)} 
\end{equation}

We insert the expression (\ref{eq:phi_i(s)}) for $\hat{\phi}_{i}\left(s\right)$ and the expression (\ref{eq:q(s)}) for $\hat{q}_{i}\left(s\right)$ into the equation for the walker density (\ref{eq:n_i(s)}) to obtain a generalization of the MW equation \cite{montroll1965} that applies to arbitrary network structures: 
\begin{align*} 
	\hat{n}_{i}\left(s\right) 
	&=\frac{1}{s}\left(1-\hat{T}_{i}\left(s\right)\right)\sum_{k}\left(I-\hat{T}\left(s\right)\right)_{ik}^{-1}n_{k}\left(0\right)\nonumber \\ 
	&=\sum_{jk}\frac{1}{s}\left(I_{ij}-\hat{T}_{i}\left(s\right)\delta_{ij}\right)\left(I-\hat{T}\left(s\right)\right)_{jk}^{-1}n_{k}\left(0\right)\,.
\end{align*} 
In terms of vectors and matrices, we write this as
\begin{align} 
	\hat{{n}}\left(s\right) 
	& =\frac{1}{s}\left(I-\hat{D}_{T}\left(s\right)\right)\left(I-\hat{T}\left(s\right)\right)^{-1}{n}\left(0\right)\,,
	\label{eq:montroll-weiss} 
\end{align} 
where the diagonal matrix $\hat{D}_{T}$ is defined as
\begin{equation*}
	\left(\hat{D}_{T}\right)_{ij}\left(s\right)\equiv\hat{T}_{i}\left(s\right)\delta_{ij}\,.
\end{equation*}
Equation (\ref{eq:montroll-weiss}) is a formal solution in Laplace space for the density of a random walk whose dynamics are governed by the WTDs $\psi_{ij}\left(t\right)$. However, taking the inverse Laplace transform to obtain the random-walker density as a function of time does not in general yield closed-form solutions.

The non-Markovian nature of equation (\ref{eq:montroll-weiss}) becomes clear after taking an inverse Laplace transform and returning to a description in the original time variable. This leads to the integro-differential equation 
\begin{equation}
	\frac{d{n}}{dt}=\left(T\left(t\right)*\mathcal{L}^{-1}\left\{\hat{D}_{T}^{-1}\left(s\right)\right\}-\delta\left(t\right)\right)*K\left(t\right)*{n}\left(t\right)\,, 
	\label{eq:master} 
\end{equation}
where $\mathcal{L}^{-1}$ denotes the inverse Laplace transform and 
\[
	f*g=\int_{0}^{t}d\tau f\left(t-\tau\right)g\left(\tau\right)
\] 
denotes convolution with respect to time. The \emph{memory kernel} $K$ characterizes the amount of memory in the dynamics \cite{balescu} and is defined in Laplace space by 
\[
	\hat{K}\left(s\right)\equiv\frac{s\hat{D}_{T}\left(s\right)}{I-\hat{D}_{T}\left(s\right)}\,.
\]
Because of the convolutions, the temporal evolution of the density of walkers at time $t$ depends on the states of the system at all times since the initial condition. 

Although the integro-differential equation (\ref{eq:master}) is instructive, it is difficult to manipulate in practice, and it is often significantly easier to analyze the associated Laplace-space equation \eqref{eq:montroll-weiss}. For example, the total number of walkers is conserved if and only if
\begin{equation*}
	\frac{d}{dt}\sum_{i}n_{i}\left(t\right)=0\,,
\end{equation*}
which is difficult to verify in the time domain. We refer the reader to our paper \cite{hoffmann} for a detailed derivation of Eq.~\eqref{eq:master} and a proof in Laplace space that the total number of walkers is indeed conserved.


\section{Asymptotic State}
\label{sec:sss}

\subsection{Preliminaries} 

Based on our analysis of random walks on static networks, we expect the walker distribution to reach a unique steady-state solution as $t \rightarrow \infty$ as long as any node can be reached from any other node. The asymptotic steady-state walker density ${p}$ satisfies
\[
	{p}=\lim_{t\rightarrow\infty}{n}\left(t\right)=\lim_{s\rightarrow0}s\hat{{n}}\left(s\right)\,,
\] 
so
\begin{align} 
	\label{eqqq}
	p & = M{n}\left(0\right)\,, 
\end{align} 
where the matrix
\[
	M\equiv\lim_{s\rightarrow0}\left(I-\hat{D}\left(s\right)\right)\left(I-\hat{T}\left(s\right)\right)^{-1}
\]
maps the initial state $n\left(0\right)$ to the final state $p$. In the limit $s \rightarrow 0$, one can expand the exponential in the definition of the Laplace transform to first order: 
\begin{align*}
	\left[\hat{D}_{T}\left(s\right)\right]_{ij}
	&=\int_{0}^{\infty}\left(1-st\right)T_{j}\left(t\right)\delta_{ij}dt+{O}\left(s^{2}\right)\\
	& =\left(1-s\langle t_{j}\rangle\right)\delta_{ij}+{O}\left(s^{2}\right)\\ 
	& =\left(I-sD_{\langle t\rangle}\right)_{ij}+{O}\left(s^{2}\right)\,, 
\end{align*} 
where the resting time 
\begin{equation}
	\label{eq:resting}
	\langle t_{j}\rangle=\int_{0}^{\infty}tT_{j}\left(t\right)dt
\end{equation}
 is the mean time spent on node $j$ and we have defined the diagonal matrix 
\begin{equation*} 
	 \left[D_{\langle t\rangle}\right]_{ij} \equiv \langle t_{j}\rangle\delta_{ij}\,. 
\end{equation*} 
 Similarly, one can use the approximation 
\begin{align}
	\label{coot} 
	\hat{T}_{ij} 
	& =\int_{0}^{\infty}\left[1-st+{O}\left(s^{2}\right)\right]T_{ij}\left(t\right)dt \notag \\ 
	& =\mathbb{T}_{ij}\left(1-s\int_{0}^{\infty}t\frac{T_{ij}\left(t\right)}{\mathbb{T}_{ij}}dt+{O}\left(s^{2}\right)\right) \notag \\ 
	& =\mathbb{T}_{ij}\left[1-s\langle t_{ij}\rangle+{O}\left(s^{2}\right)\right]\,, 
\end{align} 
where $\langle t_{ij}\rangle$ is the mean time before making a step $j\rightarrow i$ and the components of the \emph{effective transition matrix} $\mathbb{T}$ are
\begin{equation} \label{eq:effective}
	\mathbb{T}_{ij} \equiv \int_0^\infty T_{ij}(t) dt\,.
\end{equation}
We write (\ref{coot}) in matrix form as 
\[
	\hat{T} = \mathbb{T}-s\mathbb{T}\circ\langle t\rangle+{O}\left(s^{2}\right)\,, 
\]
where $\circ$ denotes the Hadamard component-wise product. This yields our final expression:
\begin{equation}\label{eq:mfinal}
	M=\lim_{s\rightarrow0}\left[sD_{\langle t\rangle}\left(I-\mathbb{T}-s\mathbb{T}\circ\langle t\rangle\right)^{-1}+{O}\left(s^{2}\right)\right]\,.
\end{equation}


\subsection{Effective Transition Matrix} \label{sec:effective}

Before focusing on the stationary state $p$ in equation (\ref{eqqq}), we discuss the properties of  $\mathbb{T}$. The matrix element $\mathbb{T}_{ij}$ is the probability of making a step $j\rightarrow i$ for any time $t \in [0,\infty)$. Thus, $\mathbb{T}_{ij} \geq 0$ for all $i$ and $j$.  Additionally,
\begin{equation}\label{eq:eff}
	\sum_i \mathbb{T}_{ij} =1\,,
\end{equation}
so $\mathbb{T}$ is a stochastic matrix (which we call the ``effective transition matrix" of the stochastic process). Because $\mathcal{G}$ is strongly connected, the stochastic matrix is irreducible and its dominant eigenvector $x$ satisfies
\begin{equation}\label{eq:domi}
	\mathbb{T} x = x\,,
\end{equation}
which has an eigenvalue of $1$ and is unique \cite{stewart2009probability}. As we will see below, the matrix $\mathbb{T}$ plays an important role in determining the asymptotic state of the system as $t \rightarrow \infty$. 

By definition, (\ref{eq:eff}) is equivalent to the condition
\begin{equation*}
	\int_{0}^{\infty}T_{j}\left(t\right)dt=1\,, 
\end{equation*}
which is expected to be true if node $j$ is connected to at least one other node.\footnote{This condition holds in our setting, as we have assumed that the underlying graph $\mathcal{G}$ of potential edges is strongly connected.} Therefore, a transition from $j$ to some other node is guaranteed to occur eventually if one allows infinite time. To show this, we use equations (\ref{eq:T_ij(t)}) and (\ref{eq:T_i(t)}) to obtain
\begin{align*} 
	T_{j}\left(t\right) =\sum_{i=1}^{N}T_{ij}\left(t\right) 
	& =-\sum_{i=1}^{N}\left(\frac{d\chi_{ij}\left(t\right)}{dt}\times\prod_{k\neq i}\chi_{kj}\left(t\right)\right)\\ 
	& =-\frac{d}{dt}\left(\prod_{i=1}^{N}\chi_{ij}\left(t\right)\right)\,.
\end{align*} 
Integrating over the entire time domain yields 
\begin{align*}
	\int_{0}^{\infty}T_{j}\left(t\right)dt 
	& =-\int_{0}^{\infty}dt\frac{d}{dt}\left(\prod_{i=1}^{N}\chi_{ij}\left(t\right)\right)\\
	& =-\left.\left(\prod_{i=1}^{N}\chi_{ij}\left(t\right)\right)\right|_{t=0}^{\infty} =1\,, 
\end{align*} 
because $\chi_{ij}\left(0\right)=1$ and $\chi_{ij}\left(\infty\right)=0$ when the edge $j\rightarrow i$ exists in the underlying graph.


\subsection{Steady-State Solutions} 

The matrix $M$ in equation (\ref{eq:mfinal}) maps any initial condition onto a unique vector as long as the underlying graph $\mathcal{G}$ consists of a single strongly connected component \cite{hoffmann}. The steady-state solution $p$ is then given by the dominant eigenvector  of the matrix $M$. In practice, it is easier to obtain the least dominant eigenvector of its inverse
\begin{align}\label{second}
	M^{-1} 
	& =\lim_{s\rightarrow0}\frac{1}{s}\left[I-\mathbb{T}-s\mathbb{T}\circ\langle t\rangle\right] D_{\langle t\rangle}^{-1} \notag \\ 
	&=\lim_{s\rightarrow0}\left[\frac{1}{s}\left(I-\mathbb{T}\right)D_{\langle t\rangle}^{-1}-\left(\mathbb{T}\circ\langle t\rangle\right)D_{\langle t\rangle}^{-1}\right]\,. 
\end{align}
In the limit $s\rightarrow 0$, the eigenvectors of $M^{-1}$ tend to the eigenvectors of the matrix 
\[
	C\equiv\left(I-\mathbb{T}\right)D_{\langle t\rangle}^{-1}\,,
\]
because the second term in the second line of (\ref{second}) becomes negligible in comparison to the first term. Thus, in the limit $s\rightarrow 0$, finding the dominant eigenvector of $M$ reduces to finding the least dominant eigenvector of $C$.  It follows that 
\begin{align*} 
	C{p} &=\left(I-\mathbb{T}\right)D_{\langle t\rangle}^{-1}D_{\langle t\rangle}{x}\\ 
& =I{x}-\mathbb{T}{x} =0\,,
\end{align*} 
where we recall that $x$ is the dominant eigenvector of $\mathbb{T}$.  It follows that the steady-state solution is
\begin{equation}\label{soloo}
	{p}=\beta D_{\langle t\rangle}{x}\,, 
\end{equation}
where $\beta$ is a normalization constant.


The equilibrium solution in (\ref{soloo}) takes a particularly simple form that is similar to the stationary solution for Markovian continuous-time random walks on static networks (\ref{solgeneral}). The time spent on node $i$ is given by the frequency to arrive on $i$ 
multiplied by the waiting time $\langle t_{i}\rangle$ spent on $i$. One can compute this solution easily even for very large graphs, because deriving $\mathbb{T}$ from $\psi$ is straightforward and one can compute the dominant eigenvector of a large matrix using standard, efficient techniques (such as the power method \cite{langville2006google}). 


\subsection{Example: Edges Governed by Poisson Processes}

We now focus on the particular case in which the underlying network $\mathcal{G}$ is undirected and the edge dynamics are governed by Poisson processes. The WTDs are then given by exponential distributions \cite{stewart2009probability} 
\begin{align}
	\psi_{ij}\left(t\right)=\lambda_{ij}e^{-\lambda_{ij}t}\,, 
	\label{lambda}
\end{align} 
where $\lambda_{ij}$ is the characteristic rate for the transition $j\rightarrow i$.  In this situation, equation (\ref{eq:T_ij(t)}) becomes
\begin{align}\label{ttt} 
	T_{ij}\left(t\right) &=  \lambda_{ij}e^{-\lambda_{ij}t}\prod_{l\neq i}\left(1-\int_{0}^{t}\lambda_{lj}e^{-\lambda_{lj}t'}dt'\right) = \lambda_{ij}e^{-\Lambda_{j}t}\,, 
\end{align} 
and equation (\ref{eq:T_i(t)}) becomes
\[
	T_{j}\left(t\right)= \sum_{i=1}^{N}\lambda_{ij}e^{-\Lambda_{j}t} =\Lambda_{j}e^{-\Lambda_{j}t}\,, 
\]
where the aggregate transition rate from node $j$ is defined as $\Lambda_{j}\equiv\sum_{i=1}^{N}\lambda_{ij}$. It follows from equation (\ref{ttt}) that the effective transition matrix $\mathbb{T}$ satisfies 
\begin{equation}\label{popopo}
	\mathbb{T}_{ij}= \frac{\lambda_{ij}}{\sum_{i=1}^{N}\lambda_{ij}}=\frac{\lambda_{ij}}{\Lambda_j}\,.
\end{equation}
The probability to follow an edge is thus proportional to its weight, and we recover the usual rate equation (\ref{ctrwgeneral}).  In particular, we obtain
\begin{equation}
	\frac{dn_{i}}{dt}=\sum_{j}\lambda_{ij}n_{j}\left(t\right) - \Lambda_{i}n_{i}\left(t\right)\,, \label{eq:rate} 
\end{equation} 
so the dynamics are governed by the combinatorial Laplacian $L_{ij}=\lambda_{ij} - \Lambda_{i} \delta_{ij}$ of a weighted network defined by the adjacency matrix $\lambda$ with components $\lambda_{ij}$. As we discussed in Section 2, the steady-state solution $p$ is thus uniform (independent of the details of the rate matrix $\lambda$).

Equation (\ref{eq:rate}) shows that a random walk on a stochastic temporal network driven by (\ref{lambda}) is equivalent to a Poisson continuous-time random walk on a static network that is constructed by aggregating a temporal network by counting the number of times edges appear between each pair of nodes. 


\subsection{Numerical Experiments}

In this section, we consider a toy example of a completely connected graph with $N = 3$ nodes to illustrate how the nature of the WTDs can affect dynamics. 

\begin{figure} 
\includegraphics[width=0.42\textwidth]{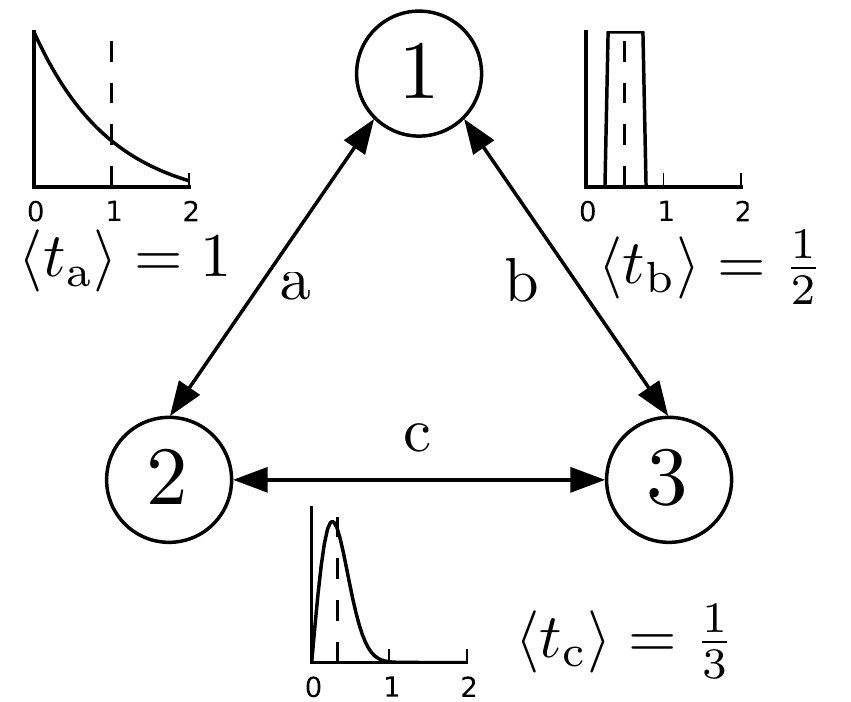}
\hspace{1cm}
	\includegraphics[width=0.42\textwidth]{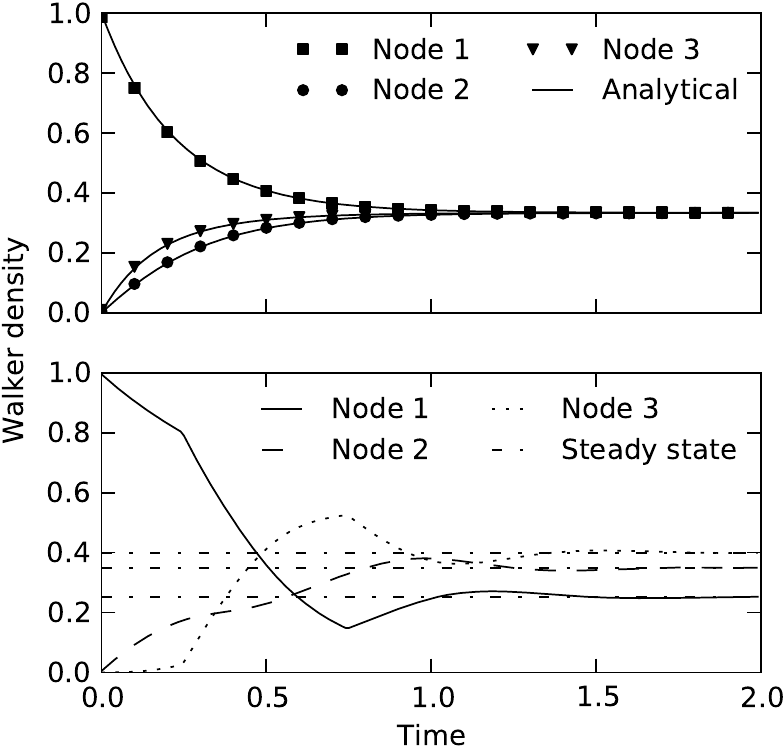}
	\caption{\label{fig:analytic-numerical} (Left) Illustration of random-walk dynamics on an undirected network with $N=3$ nodes and no self-loops. The waiting-time distributions for the edges $a$, $b$, and $c$ are exponential, uniform, and Rayleigh, respectively.  Their corresponding means are $\langle t_a\rangle$, $\langle t_b\rangle$, and $\langle t_c\rangle$. (Right) Mean temporal evolution of the random-walker densities obtained from $10^6$ numerical simulations for (upper panel) Poisson and (lower panel) non-Poisson processes. In the former case, we also plot the corresponding analytical solutions of the rate equation. The error bars are smaller than the width of the curves that we used for plotting.  For the non-Poisson example, we obtain the steady-state walker densities from equation (\ref{soloo}).  
} 
\end{figure}

We suppose that the WTDs for the processes occurring on the edges have different functional forms and different characteristic times (see Fig.~\ref{fig:analytic-numerical}), and we compare this non-Poisson situation to a situation with edges governed by Poisson processes with the same mean rates.  In our example, the mean rate matrix 
\[ 
	\lambda=\left(\begin{array}{ccc} 
	0 & 1 & 2\\ 
	1 & 0 & 3\\ 
	2 & 3 & 0 \end{array}\right)
\]
is identical in both situations. However, the resulting effective transition matrices, mean resting times, and stationary solutions differ (see Table~\ref{tbl:comparison}). We plot the temporal evolution of the walker densities in Fig.~\ref{fig:analytic-numerical} to illustrate the differences between the two situations. We obtain walker densities from numerical simulations of a random walk with all walkers located initially at node $1$. The system relaxes towards a stationary solution in both cases, but this stationary solution clearly depends on the nature of the WTDs, as walkers tend to be underrepresented on node $1$ for the non-Poisson dynamics. See our paper \cite{hoffmann} for additional numerical experiments and other details.

\begin{table}
\centering
\begin{tabular}{ll|lrl}
Symbol & Definition & Poisson & non-Poisson\\\hline\hline

Effective transition matrix ($\mathbb{T}$) & Eq.~\eqref{eq:effective} & $\left(\begin{array}{ccc}
0.000 	& 0.250 	& 0.400\\
0.333 	& 0.000 	& 0.600\\
0.667 	& 0.750 	& 0.000
\end{array}\right)$ & $\left(\begin{array}{ccc}
0.000	& 0.387 	& 0.772\\
0.273 	& 0.000	& 0.228\\
0.727 	& 0.613 	& 0.000
\end{array}\right)$\\
\hline
Dominant eigenvector ($x$) & Eq.~\eqref{eq:domi} & $\left(0.250, 0.333, 0.417\right)^T$ & $\left(0.392, 0.200, 0.408\right)^T$\\
\hline
Mean resting time ($\langle t_i\rangle$) & Eq.~\eqref{eq:resting} & $\left(0.333, 0.250, 0.200\right)^T$ & $\left(0.273, 0.387, 0.301\right)^T$\\
\hline
Stationary solution ($p=\beta D_{\langle t\rangle}x$) & Eq.~\eqref{soloo} & $\left(0.333,0.333,0.333\right)^T$& $\left(0.349, 0.252, 0.399\right)^T$\\

\end{tabular}
\caption{\label{tbl:comparison}Comparison of the effective transition matrix, dominant eigenvector, mean resting times, and stationary solution for a Poisson random walk and a non-Poisson random walk on a toy network with $N=3$ nodes. Incorrect normalizations of walker densities are due to rounding errors.}
\end{table}

\section{Discussion}

The main purpose of this chapter was to develop a mathematical framework that allows one to incorporate nontrivial temporal statistics into the study of networks. We have proposed a simple stochastic model for temporal networks and considered how stochastic processes on edges can affect diffusion processes.  In particular, we have conducted an analytical study of a random walk on stochastic temporal networks in which we demonstrated that its dynamics are driven by an integro-differential master equation. Despite the complexity of this non-Markovian process, we have found an analytical expression for its asymptotic steady-state solution and have verified its validity in numerical experiments. 

We believe that our approach offers an interesting compromise between abstract but unrealistic models and data-driven but non-mathematical approaches for studying temporal networks. However, our work does suffer from limitations that might limit its applicability in practical contexts, and attempting to remove these limitations offers several interesting research directions. First, we summarized the temporal statistics using only inter-event time distributions and thereby neglected higher-order temporal correlations \cite{karsai2}. Second, we examined stationary-state dynamics of temporal networks, and many systems that can be modeled using temporal networks do not attain stationary states (due, for example, to daily or weekly patterns). Finally, we have implicitly assumed when solving the generalized Montroll-Weiss equation (\ref{eq:montroll-weiss}) that the first moments of $\psi_{ij}(t)$ and $T_{ij}(t)$ are finite. In Section 5.1, we performed a small-$s$ expansion and used the quantities $\langle t_{j}\rangle$ and $\langle t_{ij}\rangle$ in our calculations. These operations are not valid when it is not possible to define a characteristic waiting time between two steps.  (This occurs, for example, when the Laplace transform of the WTDs behaves near $s = 0$ like $1 - a s^\alpha$, where $\alpha<1$ \cite{balescu}.) Because empirical observations of WTDs in real-world systems often have heavy tails \cite{bara}, it would be interesting to extend our results to this situation.

We expect our approach to pave the way for the development of new tools that consider both network structure and network dynamics \cite{lambiotte}. For example, there is an urgent need for the development of approaches to analyze networks that properly take into account the temporal dynamics of edges \cite{temporalmetrics,grindrod,pathlengthtemporal,Mucha}. A direct application of our work is to define a modified version of PageRank centrality (which is sometimes called simply PageRank) \cite{pagerank}, as well as modified versions of related measures \cite{review}, to measure the importance of nodes with respect to the dynamics of a random walker on a temporal network. PageRank is a conservative \cite{lerman}, non-local measure of node \emph{centrality} (i.e., of a node's importance \cite{review}) that has been applied to a huge variety of networks across a broad range of scholarly disciplines both in its original form \cite{langville2006google,bergstrom2008eigenfactor,radicchi2011best,review} and in slight variations of it \cite{Callaghan,allesina2009googling,lamros}. The PageRank vector is usually defined for discrete-time random walks, and the component of this vector corresponding to a specific node is given by the expected density of random walkers on that node at stationarity (i.e., by the frequency at which that node is visited in the time $t \rightarrow \infty$ limit). The PageRank vector is equal to the dominant eigenvector of the transition matrix $T$, whose components are given by (\ref{unbiased}). The case of PageRank for a continuous-time process is somewhat more complicated, as the density of walkers $p$ and frequency of visits $x$ are now different in general. As we have discussed in this chapter, these two quantities are related by the relation ${p}=\beta D_{\langle t\rangle}{x}$ at stationarity, where $D_{\langle t\rangle}$ is the mean time spent on a node before leaving it. To ensure that the standard PageRank vector is recovered in the Poisson limit, we choose to use $x$ for the PageRank centrality for continuous-time processes. Accordingly, we define PageRank on stochastic temporal networks as the dominant eigenvector of the effective transition matrix $\mathbb{T}$, as this gives more importance to edges that are visited more by walkers due to the temporal order of their appearance. It would be interesting to explore the properties of this centrality measure and to compare it to PageRank and other existing (conservative) notions of centrality.  It would also be interesting to develop non-conservative centrality notions for temporal networks, and one has to consider processes other than random walks to do that \cite{lerman}. Other possible applications of our work include the construction of random-walk based measures of network modularity \cite{rosvall,delvenne,Mucha} or node similarity \cite{jeh} for stochastic temporal networks.


\begin{acknowledgement}

This chapter is based on Ref.~\cite{hoffmann}, which contains additional calculations and numerical simulations. RL would like to acknowledge support from FNRS (MIS-2012-F.4527.12) and Belspo (PAI Dysco). MAP acknowledges a research award (\#220020177) from the James S. McDonnell Foundation and a grant from the EPSRC (EP/J001759/1).

\end{acknowledgement}

\input{referenc2}

\end{document}

%% file: referenc2.tex
%
%
%